\documentclass{article}[12pt]
\usepackage{amsmath}
\usepackage{amssymb}
\usepackage{amsthm}
\usepackage{amsfonts}
\usepackage[dvips]{graphicx}

\def\abstract#1{\vskip 7mm
        \begin{center}{\large Abstract}\par \smallskip
                \begin{minipage}[c]{12cm}
                        \small #1
                \end{minipage}
        \end{center}
}

\def\title#1{\begin{center}{\Large\bf #1}\end{center}}
\def\author#1{\vskip 5mm \begin{center}{#1}\end{center}}
\def\address#1{\begin{center}{\it #1}\end{center}}

\newcommand{\ssmatrix}[4]%
{\begin{pmatrix} #1 & #2 \\ #3 & #4 \end{pmatrix}}
\makeatletter
\@ifundefined{lesssim}{}{}
\@ifundefined{gtrsim}{}{}
\def\vereq#1#2{\lower3pt\vbox{\baselineskip1.5pt \lineskip1.5pt
\ialign{$\m@th#1\hfill##\hfil$\crcr#2\crcr\sim\crcr}}}
\makeatother

\newtheorem{definition}{Definition}

\newtheorem{lemma}{Lemma}
\newtheorem{proposition}{Proposition}

\newcommand{\MMI}{Main Theorem I}
\newcommand{\MMII}{Main Theorem II}

\theoremstyle{definition}
\theoremstyle{remark}
\newcommand{\vtr}[2]{\paren{\begin{array}{c} #1 \\ #2 \end{array}}}

\newcommand{\EPS}{{\cal E}}
\newcommand{\eh}{event horizon}
\newcommand{\nn}{\nonumber \\ }
\newcommand{\spt}{spacetime}
\newcommand{\ie}{{\em i.e.\/}}
\newcommand{\Mset}{B_{\rm Maxwell}}

\newcommand{\crset}{{\cal C}}
\newcommand{\diffset}{{\cal D}}
\newcommand{\R}{{\mathbb R}}

\newcommand{\MM}{{\cal M}}
\newcommand{\HH}{{\cal H}}
\newcommand{\CS}{{\cal S}}
\newcommand{\UU}{{\cal U}}
\newcommand{\bfem}{\em}
\newcommand{\Cinf}{\mbox{$C^\infty$}}
\newcommand{\paren}[1]{\left({#1}\right)}

\newcommand{\ug}{unfolding}

\newcommand{\ugs}{unfoldings}

\newcommand{\Ugs}{Unfoldings}
\newcommand{\WT}{Whitney {\Cinf} topology}
\newcommand{\Ref}[1]{(\ref{#1})}
\newcommand{\req}{\simeq}
\newcommand{\minimal}{minimum}
\newcommand{\nbhd}{neighbourhood}
\newcommand{\mf}{\mu_F}

\begin{document}
\setlength{\baselineskip}{16pt}
\title{%
Topology of event horizon in axially symmetric spacetime:
Classification of Maxwell set with symmetry
}
\author{%
  Masaru Siino\footnote{E-mail:msiino@th.phys.titech.ac.jp}
}
\address{%
  Department of Physics, Tokyo Institute of Technology, \\
  Oh-Okayama, Megro-ku, Tokyo 152-, Japan
}

\abstract{
The crease set of an event horizon is studied in a spacetime with discrete or
continuous symmetry. 
It determines possible topologies on spatial sections of an event horizon.
We thereby investigate the classification of stable topological structure of the crease sets in a spacetime with any symmetry. 
In practice, we show the classification of the crease set in axially symmetric spacetime. 
By that we realize the topological structure of axially symmetric event horizons.
We will finc that many new topological structures become stable which is not stable without symmetry.}

\section{Introduction}
The final state of the black hole is well understood and is simply described by at most three parameters.
On the other hand, an early stage of black hole formation is not well understood since there is much variety of their appearance.

In the topological viewpoint, the final state was investigated by many authors and now it is known that under some reasonable conditions such as asymptotic flatness and the weak energy condition, each component of the black hole region is topologically trivial, {\ie} simply connected\cite{OW}. 
On the other hand, there were numerical simulations which suggest non-trivial topologies of event horizons in an early stage of black hole formation\cite{ST}\cite{NCSA}. 
There has been some confusion, but it is now well understood that 
even though the black hole region in the {\spt} is simply connected, 
there are many possible topologies of spatial sections. 

In particular, it is revealed that various topologies of the spatial sections of black hole are determined by the endpoint set, or similarly, the {\em crease set}, of the event horizon by the author\cite{MS1}.
Therefore the  crease set of an event horizon, is an important object which is independent from the choice of time slices and which determines qualitative properties of the event horizon.
Thus, in order to restrict the physically possible topologies of the black hole, it is important to restrict the possible (stable) structure of the crease set. 
Since the crease set can be regarded as a singularity of a distant function (i.e. the Maxwell set) determining the generators of the event horizon. 
The singularity theory of real maps can classify the stable topological structure of the crease set locally\cite{MS2}\cite{SK}\cite{AN}.

These possible topological structures should be studied in numerical experiments of gravitational collapse\cite{ST}\cite{NCSA}\cite{HW}. Nevertheless, it might be very hard since it requires a huge data set of a geometry in the entirety of numerically generated spacetime, to determine its event horizon.
To reduce the cost of the numerical calculation, it may be successful to impose any symmetry on a black hole spacetime. 
Indeed it is in axially symmetric gravitational collapse that some remarkable results are reported about non-trivial topologies of the event horizon\cite{ST}\cite{NCSA}.

In the symmetric spacetime, however, is the previous result\cite{SK} how to be applied to?
In a rigorously symmetric spacetime, their perturbation should be also symmetric.
Then the stability of the function determining the generators of an event horizon will be altered. 
We need to study the classifications of the crease set again in this symmetric situation.
By this, some new crease sets are added to the classification without symmetry.

This investigation will become relevant to also an almost symmetric realistic spacetime, if the symmetry is
dynamically stable and if non-symmetric perturbation is always sufficiently small.
In such a situation, the crease set in the almost symmetric spacetime will be approximated
by the symmetric classification of the crease set in rigorously symmetric spacetime, by neglecting 
very fine structures of the crease set (c.f. \cite{CP}). 

In the present article, we investigate the classification of the Maxwell set under discrete and continuous (especially, reflection and axial) symmetry. 
Of course, the classification includes the event horizons reported in references \cite{ST}\cite{NCSA}.

In the next section, we mention the definition of variational principle to determine the null generators of event horizons in symmetric spacetime. 
The crease set is mathematically related to the Maxwell set of a potential function in the third section. 
We investigate the classification of the symmetric Maxwell set in the fourth section. 
The fifth section gives a concrete classification in reflection symmetric three-dimensional spacetime and  axially symmetric four-dimensional spacetime. 
Then we will discuss the spatial topology of axially symmetric black hole.
The final section is devoted to summary and discussions.

\section{Fermat potential in symmetric spacetime}

An event horizon is generated by null geodesics. 
A future {\eh} $\HH$ 
cannot have future endpoints, but 
can have past endpoints if it is not eternal. 
As is pointed out in \cite{MS1}, 
the {\em endpoint set}\ $\EPS$ of a horizon 
is an arc-wise connected acausal set. 
Points $u\in\EPS$ are classified by 
the {\em multiplicity\/} $m(u)$ of $u$, 
the number of the generators emanating from $u$:
\begin{eqnarray}
  \crset&:=&\{u\in\EPS\;|\;m(u)>1\},\nn
  \diffset&:=&\{u\in\EPS\;|\;m(u)=1\}.
\end{eqnarray}
The set $\crset$ is called the {\em crease set} of the horizon. 
The crease set contains the interior of the endpoint set, 
{\ie},
the closure of $\crset$ contains $\EPS$~\cite{BK}. 
The crease set $\crset$ equals the set of points of $\EPS$ on which the horizon is not differentiable, {\ie},  the horizon is differentiable at $u\in\EPS$ if and only if $u\in\diffset$~\cite{BK,Chr98CQG}.

The horizon $\HH$ is the envelope of the light cone starting from the
crease set $\crset$ 
which is an arc-wise connected acausal subset of $\HH$. 
If the spatial section of the horizon 
is a topological sphere at late times, 
the topology of the spatial section of the horizon can be nontrivial
only at the crease set and the topology is completely determined by
the time slicing of the crease  set, which is studied in Ref.~\cite{MS1} in detail. 
In particular, when the crease set
is a single point, each possible spatial section of the horizon topologically is a single sphere. 
On the contrary, when the crease set has a disk-like structure, the horizon can have toroidal or higher-genus spatial sections. 
One would see the coalesce of horizons if the crease set has a line-like structure\cite{NCSA}. 
Therefore by classifying the structure of the crease set, we will know all possible topologies of the horizons. 
Here we do not assume that the spatial section of the horizon in the future is a sphere. 

The crease set can be 
determined by Fermat's principle in simple stationary {\spt}\cite{MS2}. 
In non-stationary {\spt}, 
we can extend Fermat's principle 
and find a variational principle about light paths, 
imposing some appropriate causality condition such as 
global hyperbolicity. 
Here we show an example of the construction of the Fermat potential, and see how the symmetry of spacetime impose a condition to the Fermat potential.

Let us assume that the {\spt} $\MM$ is smooth and is 
globally hyperbolic from 
a smooth Cauchy surface $\CS$ which is diffeomorphic to $\R^n$. 
Furthermore, we consider {\spt} of gravitational collapse, 
namely, we assume that the event horizon $\HH$ is in the future of
$\CS$. 

By global hyperbolicity, there are
always an appropriate smooth global time coordinate $t:\MM\to\R$ and a 
timelike vector field $T$ such that $dt(T)=1$. 
The {\spt} $\MM$ is foliated by Cauchy surfaces 
${\CS}_t=\{q\in\MM|t(q)=t\}$. 
The vector field $T=\partial/\partial t$ 
defines a smooth projection $\pi$ from $\MM$ into the
$\CS={\CS}_{t_0}$; 
\begin{align}
  &\pi:{\cal M}\to {\cal S},\nn
  &  \pi^{-1}(q)=\{ \gamma(t) |
  \frac{\partial \gamma}{\partial t}=T,
  \gamma(t_0)=q, t\in\R
  \}. 
\end{align}
Conversely, there is a diffeomorphism
\begin{align}
  &\phi:\R\times\CS\to\MM, \nn
  &t(\phi(t,u))=t, \quad \pi(\phi(t,u))=u. 
\end{align}
Because $\HH$ is achronal, the restriction of $\pi$ on $\HH$ 
is injective and has an inverse, which we denote by $\psi$:
\begin{align}
  &\psi:\pi(\HH)\to\HH, \nn
  &\psi(u)\in\HH,\quad \pi(\psi(u))=u. 
\end{align}
The map $\psi$ is Lipschitz\cite{HE}.


In the present article, we suppose that the spacetime possesses any symmetry.
This means that there exists a symmetric Cauchy surfaces ${\cal S}_{St}$ and there are any
 isometries $\varphi$ acting on each spatial hypersurface ${\cal S}_{St}$ isometrically,
\begin{align}
\varphi:(\MM,g)&\mapsto (\MM,g) \\
s.t.\ \varphi({\cal S}_{St})&={\cal S}_{St}\\
\varphi^*g=g&, \varphi^*h=h,
\label{eqn:isom}
\end{align}
where $h$ is a spatially induced metric on ${\cal S}_{St}$.
We consider both cases of discrete symmetry and continuous symmetry.
For these cases their isometries form discrete group and continuous group, respectively.

Then we adopt a time function of this symmetric timeslicing as a symmetric global time coordinate 
$t(q)$ and its normal time vector $T=\partial/\partial t\perp {\cal S}_{St}$ defines symmetric smooth projection $\pi$.
By the fact that $\partial/\partial t$ and the Killing vector generating the isometry (\ref{eqn:isom}) are orthogonal, $t$
can be a synchronous time,
\begin{equation}
g\left(\frac{\partial}{\partial t},\frac{\partial}{\partial t}\right)=const.
\end{equation}


We take some (sufficiently large) $t=t_1$ and assume that 
$\CS_{St_1}\cap\HH$ is diffeomorphic to a compact manifold $M$. 
We consider $M$ as a fixed submanifold embedded in $\CS_S$ so that 
$\CS_{St_1}\cap\HH=\psi(M)$. 
Consider a {\nbhd} $U$ of $\pi(\HH\cap J^-(\CS_{St_1}))$ in $\CS_S$.
For $x\in M$ and $u\in U$ we define Fermat
potential as follows:
\begin{align}
  F(x,u):=-\sup \{t\in\R|\phi(t,u)\in J^-(\psi(x))\}.
  \label{eq-fermat}
\end{align}

Obviously if there is an event horizon, it should be invariant under the isometry $\varphi$
\begin{equation}
\varphi(\HH)=\HH.
\label{eqn:invH}
\end{equation}
Then $M$ is also invariant under the isometry $\varphi$
\begin{equation}
\varphi(M)=M.
\label{eqn:invM}
\end{equation}
Furthermore we take such $U$ that it is also invariant subset, $\varphi(U)=U$.

The minimum points $x$ of $F$ corresponds to the generator of $\HH$ through $u$. 
Our definition \Ref{eq-fermat} is the generalization of the geodesic distance function to the non-static {\spt}.


From the above construction of the Fermat potential,
it will be understood how the Fermat
potential reflects the symmetry of spacetime.
Since we adopt a spatially symmetric timeslicing ${\cal S}_{S}$ and its synchronous normal time vector ${\partial}/{\partial t}$,
$F(x,u)$ is invariant under isometric transformation.
In this construction the isometry acts on both state $X$ and control $U$ spaces.
Then pull back of the Fermat potential satisfies

\begin{equation}
\varphi^*F(x,u)=F(\varphi(x),\varphi(u))=F(x,u), 
\label{eqn:smoF}
\end{equation}
where the isometry $\varphi$ acts on $M$ and $U$ as a invariant subset of $\cal M$ from (\ref{eqn:invH}),(\ref{eqn:invM}) and
\begin{equation}
\pi\circ\varphi=\varphi\circ\pi.
\end{equation}

From \Ref{eq-fermat}, the crease set $\crset$ is given by 
\begin{align}
  \crset=\psi(\Mset(F)),
\end{align}
where $\Mset(F)$ is the Maxwell set of $F$ where $F$ has two or more
minimum points. 
A precise mathematical framework to study its properties is given in \cite{SK}.
By (\ref{eqn:smoF}), $\cal C$ is also invariant $\varphi({\cal C})={\cal C}$.

The goal of our present study is to classify  all generally possible structure about the singularities (Maxwell set) of the Fermat potential determining the event horizon. 
As shown in ref.\cite{SK}, the generic structure will be given by studying
singularities of stable Fermat 
 potentials concretely. Then
our main task is to give a classification of the stable functions satisfying (\ref{eqn:smoF}).
When one usually discusses the bifurcation
structure, 
so-called caustics, of a system we analyze the Fermat potential locally in the context of a function germ. 
However, our main object here is not 
the bifurcation set but the Maxwell set (the difference is clarified in ref.\cite{SK}). 
The problem is not purely local but is 
rather semi-local. The
definition 
of Maxwell set is local in control (parameter) space $U$ but
is non-local in state (variable) space $M$. 
To treat this, one introduces function multigerms and classifies stable multiunfoldings and their Maxwell sets\cite{SK}. 
On the other hand, our Fermat potential in the present study is restricted to a symmetric one (\ref{eqn:smoF}).
To give a symmetric multiunfolding, non-local $U$-variable is prepared as an orbit of the isometry and also non-local $M$-variable to include non-isolated critical points of a symmetric Fermat potential.
In the next section, we give these concepts as a $G$-orbital multigerm of unfolding.

\section{mathematical preparation}
\label{maxwell}

The mathematical framework to study the Maxwell set was given in \cite{SK}.
There the semi-local characteristic of a Fermat potential is investigated in the context of a multigerm.
Nevertheless it needs some additional preparation to study the Maxwell set with any symmetry.
It is caused mainly by the change of a function space of perturbation.
That requires the extension of a multigerm so as to represent an isometry orbit of a multigerm
and competence of critical points mapped by the isometries.

First of all, we prepare the same notions as in \cite{SK}. 
A function unfolding $F:M\times U\to {\R}$ can be considered as 
a family of functions $f_u:M\to\R$ with $f_u=F(\bullet,u)$.
The Maxwell set of a function unfolding is the set of all values of the parameters for which the minimum is attained either at a non-Morse critical point or at two or more critical points.

\begin{definition}[Maxwell set]
For a function unfolding $F:M\times U\to {\R}$ on a compact
manifold $M$, 
the {\bfem Maxwell set} ${\Mset}(F)$ of $F$ is 
a subset of $U$ given by
\begin{eqnarray}
{\Mset}(F)&:=&\{u\in U| \text{$f_u$ has two or more }
\text{global {\minimal}  
  points}\}. 
\end{eqnarray}
\end{definition}

In the following we sometimes make $x$ as a
representative of the state variables and $u$ the control variables.
Of course, $F$ is not a global function on ${\R}^n\times {\R}^r$
rather a function 
on manifold $M\times U$, where $M$ is a compact manifold
and $U$ is an open subset of ${\R}^r$. 

In the investigation of the Maxwell set we mainly focus on its local structure because the global structure is obtained by the combinations of local ones.
Below we extensively use the notion of the germs of objects 
which provides the best way to characterize their local structure. 
The rigorous definition of germ is well known and will be given elsewhere\cite{SK}.
Let $M$, $N$ be \Cinf-manifolds. 
We denote the set of $\Cinf$-maps from $M$ to $N$ by $\Cinf(M,N)$.

The definition of the Maxwell set requires the global information of the function unfolding $F$. 
A simple but crucial observation is, however, that 
to determine the local structure of the Maxwell set, \ie, 
the Maxwell set germs, 
we only need the {\em local information of $F$ around its  global {\minimal}  points\/} $p_1,...,p_k\in M$. 
We generalize the notion of germs to that of multigerms.  
Let $M^{(k)}$ be a $k$-tuple of distinct points of $M$, \ie,
\begin{equation}
M^{(k)}:=\{(x_1,...,x_k)\in M^k|x_i\neq x_j\text{ for }i\neq j\}. 
\end{equation}
\begin{definition}[Multigerm]
  Let $(x_1,...,x_k)\in M^{(k)}$. 
  A {\em $k$-fold map germ\/} 
  $f:(M,(x_1,...,x_k))\to N$,
  or 
  $[f]_{x_1,...,x_k}$,
  is the equivalence class of $f\in\Cinf(M,N)$, where two maps are equivalent if they coincide on some open subset of $M$ which contains $x_1,...,x_k$. 
  A {\em $k$-fold unfolding germ\/} 
  $F:(M\times U,((x_1,...,x_k),u_0))\to N$, 
  or $[F]_{(x_1,...,x_k),u_0}$, 
  is the equivalence class of 
  $F\in\Cinf(M,\R)$ where two functions are equivalent if they coincide 
  on some open subset of $M\times U$ which contains $(x_1,u_0),...,(x_k,u_0)$. 
  A $k$-fold germ is also called as a {\em multigerm}. 
\end{definition}
A multigerm can be considered as $k$-tuple of simple germs. 
For example, a map multigerm $[F]_{(x_1,...,x_k),u_0}$ 
can be considered as $k$-tuple of function germs $([f_1]_{x_1},...,[f_k]_{x_k})$.

To study the Maxwell set under any symmetry,
first we prepare a set of multigerms they are generated by a transformation group.

\begin{definition}[$G$-orbital multigerm of unfolding]
Consider a group of transformation $G$.
  Let $(x_1,...,x_k)\in M^{(k)}$. \
$G(((x_1,...,x_k),u_0))=\{g(((x_1,...,x_k),u_0))|g\in G\}\subset M^{(k)}\times U$ is the orbit of $G$ through $((x_1,...,x_k),u_0)$.
  A {\em $G$-orbital multigerm of unfolding\/}
  $F_G:(M\times U,(G((x_1,...,x_k),u_0)))\to N$, 
  or $[F_G]_{G((x_1,...,x_k),u_0)}$, 
  is the equivalence class of $F\in\Cinf(M\times U,\R)$ where two functions are equivalent if they coincide on some open subset of $M^{(k)}\times U$ which contains $G(((x_1,...,x_k),u_0))$. 
\label{def:gom}
\end{definition}
A $G$-orbital multigerm can be considered as a suite of multigerms. 
For example, if $G$ is a discrete group a $G$-orbital multigerm $[F_G]_{G((x_1,...,x_k),u_0)}$ 
can be considered as multituple of k-tuple function germs 
$(([f_1^1]_{x_1},...,[f_k^1]_{x_k}),([f_1^g]_{g(x_1)},...,[f_k^g]_{g(x_k)}),(...),...)$ where  $g\in G$. 

If a representative of $F_G$ is invariant under actions of $g\in G$,
by definition its $G$-orbital multigerm of unfolding is invariant under actions of $G$.
Here it should be noticed that discrete $G$ gives point-wise equivalence similar to usual multigerm
otherwise the equivalence is not point-wise.
It is also commented that when $x_i$ or $u_0$ is a fixed point of $G$ the equivalence becomes
point-wise there even if $G$ is continuous.

It is required to define not point-wise equivalence of maps to
discuss the diffeomorphism class of this $G$-orbital multigerm of unfolding.

\begin{definition}[Map germ at subset]
Subset $S\subset M$ is given.
Maps $f,g\in\Cinf(M, N)$ 
are {\bfem equivalent} 
at subset $S$ if there is a {\nbhd}
$W$ of $S$ such that $f|_{W}=g|_{W}$. 
A {\bfem map germ\/} $f$ at $S$, $[f]_{S}$,  
is the equivalence class 
of $f$.
It is also denoted by 
$f:(M,S)\to N$ or 
$f:(M,S)\to(N,f(S))$. 
\end{definition}
Henceforth we call germ at subset simply germ and sometimes also germ is omitted.
Examples of map germs at subset include function germs and diffeomorphism
germs. 

As easily seen, that $G$-orbital multigerm of unfolding is not local also in $U$.
When we find the Maxwell set of it, it will be expressed as a non-local
equivalence class of set.

\begin{definition}[Set germ at subset]
Subset $S\subset U$ is given.
Subsets $X$, $Y$ of $U$ 
are {\bfem equivalent at subset $S$\/}  if 
$X\cap W=Y\cap W$ for each {\nbhd} $W$ of $S$. 
A {\bfem set germ\/} $(X,S)$ of $X$ at $S$ is the equivalence class 
of $X$. 
Set germs $(X,S)$ and $(Y,S')$ are {\bfem diffeomorphic\/}, 
$(X,S)\simeq(Y,S')$,  
if there is a diffeomorphism germ $\phi:(M,S)\to M$ 
such that 
$(\phi(X),\phi(S))=(Y,S')$.
\end{definition}

Next we give a fundamental concepts to investigate the stability of unfolding under symmetry\cite{A}\cite{KTH}.

\begin{definition}[Symmetry preserving diffeomorphism]
Let $(\MM,g)$ be a symmetric space with isometry group $G$.
A diffeomorphism $\phi$ on $(\MM,g)$ is a symmetry preserving diffeomorphism (SPD) if it keep the action $a\in G$
on $(\MM,g)$ isometry.
If $a$ diffeomorphism $\phi$ is an SPD then $\overline{\phi}:a\mapsto \phi\circ a\circ\phi^{-1}$ is an automorphism
of G.
In the following, $G$ mainly acts on invariant subset $M\times U$ as
\begin{align}
\varphi:M\times U&\mapsto M\times U\\
(x,u)\in M\times U&, \varphi(x,u)=(\varphi(x),\varphi(u)).
\end{align}
Then SPD is sometimes considered as the diffeomorphism on $M\times U$ besides $M$ and $U$ satisfying above condition.
\end{definition}

In the present article we discuss the stability defined by this SPD, since the subset of the function space we treat is invariant
under the SPD.
\begin{definition}[Right SPD equivalence]
Function germs 
$f:(M,x)\to {\R}$ and 
$g:(M,y)\to {\R}$ 
are {\bfem right equivalent\/}, 
$[f]_{x}\req [g]_{y}$, 
if there exists a SPD germ 
$\phi:(M,x)\to(M,y)$
and $a\in\R$ 
such that $f=g\circ \phi+a$ holds as an equality of 
function germs at $x$. 
\end{definition}

\begin{definition}[Right SPD equivalence at the {\minimal} points]
{\Ugs} $F:(M\times U,M\times\{u_0\})\to {\R}$ and 
$G:(M\times U,M\times\{v_0\})\to {\R}$ 
are {\bfem right equivalent at the  {\minimal} 
points\/}

if the following conditions hold: 

(1) The functions $f_{u_0}=F(\bullet,u_0)$ and 
$g_{v_0}=G(\bullet,v_0)$ have
the same number of global {\minimal}  points, 
$p_1,...,p_k$ and $q_1,...,q_k$, respectively. 

(2) 
There exists 
  a SPD multigerm
  $
    \phi:(M \times U,((p_1,...,p_k),u_0))
  \to
  (M \times U,((q_1,...,p_k),v_0))
  $, 
a SPD germ 
  $
    \psi:(U,u_0)\to(U,v_0)
  $, 
  and
  a function germ
  $\alpha:(U,u_0)\to\R$
  such that
\begin{equation}
  \label{eq-sta}
  F(x,u)=G(\phi(x,u),\psi(u))+\alpha(u)
\end{equation}
holds with both sides being 
function multigerms at $((p_1,...,p_k),u_0)$. 
\end{definition}


The Maxwell set germ of an {\ug} 
is determined only by the 
unfolding multigerm at the  {\minimal}  points: 
\begin{proposition}
If {\ugs} 
$F:(M\times U,M\times\{u_0\})\to\R$ 
and
$G:(M\times U,M\times\{v_0\})\to\R$ 
are right SPD equivalent at the {\minimal} points, then their Maxwell set germs are equivalent under SPD transformation. 
\label{col:1}
\end{proposition}
\begin{proof}
  Follows directly from 
  the definitions of right SPD equivalence and 
  of Maxwell set germs. 
\end{proof}


Below, we will determine the topological structure of $\Cinf(M,N)$
where all the maps that we treat are included. We define a topology of
$\Cinf(M,N)$ by the $r$-jet space $J^r(M,N)$ below.  

\begin{definition}[Jet space]
Let $f\in \Cinf(M,N)$. 
The {\bfem $r$-jet} $j^rf(x_0)$ of $f$ 
at $x_0$ is the equivalence class 
of $f$ in $\Cinf(M,N)$
where 
two maps are equivalent if
all of their $s$-th partial derivatives with $1\le s\le r$,
in some coordinate systems of $M$ and $N$, coincide. 
The {\bfem $r$-jet space\/} of $\Cinf(M,N)$
is defined by 
\begin{equation}
J^r(M,N):=\{j^rf(x_0)|f\in \Cinf(M,N)\}. 
\end{equation}
\end{definition}
The space of $r$-jets at a point is 
an $n\vtr{m+r}r$-dimensional manifold, 
where $m=\dim M$ and $n=\dim N$. 


Now we endow the space $\Cinf(M,N)$ with the {\WT}. 
\begin{definition}[{\WT}]
For an open subset $O$ of $J^r(M,N)$, let
\begin{equation}
W^r(O):=\{f\in \Cinf(M,N)|j^rf(M)\subset O\}.
\end{equation}
The {\bfem Whitney \Cinf topology} on $\Cinf(M,N)$ 
is the topology whose basis is 
\begin{eqnarray}
  \bigcup_{r=0}^\infty 
  \{W^r(O)|
  \text{$O$ is an open subset of $J^r(M,N)$}\}.
\end{eqnarray}

\end{definition} 
Hereafter we treat that $\Cinf(M,N)$ as a topological space with 
the {\WT}. 
Now we can define stability of the Maxwell set 
using this topology.

The concepts of the jet space and the Whitney $\Cinf$ topology are used as in the ref.\cite{SK}.
Especially, multiunfolding was investigated by the jet space for $\Cinf(M\times U,\R)$.
In the present article, however, it is not this $\Cinf(M\times U,\R)$ in which we investigate the stability of a $G$-orbital 
multigerm of unfolding but
its subset whose elements are invariant under an isometry group $G$ (\ref{eqn:smoF}).
We call it $G$-$\Cinf(M\times U,\R)$. From (\ref{eqn:smoF}), it can be identified with $C^{\infty}(\frac{M\times U}{G},
\R)$.
Here we propose that
the topology of the subset is relative topology and  equivalent to the topology of $C^{\infty}(\frac{M\times U}{G},
\R)$ in the context of the Whitney topology. $G$-$\Cinf(M\times U,\R)$ is invariant subset under the
pull back of SPD.

Now, we have completed the preparation. 
We will investigate stable structure of the function unfolding and its
Maxwell set.

\begin{definition}[$G$-stable Maxwell set germ] 
A $G$-orbital multigerm of {\ug}, 
\begin{equation}
F_G:(M\times U,G(((x_1,...,x_k),u_0)))\to {\R}, 
\end{equation}
is {\bfem stable with respect to the Maxwell set\/} 
if for each {\nbhd} $W$ of
$G(u_0)$ there exists a {\nbhd} ${\cal U}$ of $F_G$ in
$G$-$\Cinf(M\times U,{\R})$ such that for each $F_G'\in {\cal U}$ there
exists $v_0\in W$ 
such that $(\Mset(F_G),G(u_0))$ and $({\Mset}(F_G'),G(v_0))$ are symmetry preserving
diffeomorphic about $G$.

We call $({\Mset}(F),G(u_0))$ a {\bfem
  $G$-stable Maxwell set germ}. 
\label{def:gsms}
\end{definition}

\begin{definition}[$G$-stability at the minimum points]
A $G$-orbital multigerm of {\ug} 
$F_G:(M\times U, ((x_1,...,x_k),u_0))\to{\R}$
is {\bfem $G$-stable at the  {\minimal}  points} 
if
for each {\nbhd} $W$ of $G(u_0)$ 
there exists a {\nbhd} ${\cal U}$ of $F_G$ in 
$G$-$\Cinf(M\times U,{\R}$) 
such that for each $F_G'\in{\cal U}$ there exists $G(v_0)\in W$ 
such that 
$[F]_{G(u_0)}$  and $[F']_{G(v_0)}$ are right SPD equivalent at the minimum points. 
\label{def:gsmp}
\end{definition}
From Proposition~\ref{col:1}, we immediately 
have the following proposition. 
\begin{proposition}
If a $G$-orbital multigerm of {\ug} $F_G:(M\times U, M\times G(u_0))\to{\R}$
is $G$-stable at the  {\minimal}  points then 
$({\Mset}(F),G(u_0))$ is a $G$-stable Maxwell set germ. 
\end{proposition}
In this sense, our aim is to classify the $G$-stable Maxwell set germs 
of the Fermat potential $F$. 
In the present article, we will not prove this stability in practice.
We simply suggest the stability from the result of Ref.\cite{SK} without any symmetry.

To discuss the stability and classification, we study the orbit of diffeomorphism on some
standard functions in $J^r(M,\R)$ and its transversality. 
We will stratify $J^r(M,\R)$, {\ie}, 
decompose the $J^r(M,\R)$ into the union of submanifolds 
(strata)~\cite{STF}.  

\begin{definition}[Strata]
\begin{align}
A_0&:=\{j^rf(p)\in J^r(M,{\R})\,|\, 
\text{$f$ is regular at $p$}\},\\
A_k&:=\{j^rf(p)\in J^r(M,{\R})\,|\,
[f]_p\req
[\pm x_1^{k+1}]_0\}.
\end{align}
\label{def:str}
\end{definition}
The following is well discussed~\cite{STF}:
\begin{lemma}
  (1) $A_0$ is an open subset of $J^r(M,{\R})$ 
  hence is a submanifold of 
  codimension 0. 

(2) $A_k$ is a submanifold of $ J^r(M,{\R})$ of codimension $k$.

(3)  $\Sigma :=J^r(M,{\R}) -\bigcup_{k=0}^3A_k $ 
  is the union of a finite number of submanifolds 
  of codimension $4$ or greater:
\begin{equation}
\Sigma=W_1\cup ...\cup W_s,
\end{equation}
where the codimension is determined for the case $M$ is one-dimensional.
It is different from that of Ref.\cite{SK}.
\end{lemma}

\begin{definition}[Natural stratification]
The {\bfem natural stratification\/} of $J^r(M, {\R})$, where $r>3$, 
is the one given by
\begin{align}
{\cal S}(J^r(M,{\R}))
:=\{A_0,A_1,...,A_3,W_1,...,W_s \}.
\end{align}
\end{definition}
This natural stratification  corresponds to the classification of the
stable unfolding under the aid of well-known transversality
theorem.
To discuss the Maxwell set we naturally extend this concept to the one at 
several minimum points.
For that, we extend the jet space to multijet space.
\begin{definition}[Multijet space]
The {\em $k$-fold  $r$-jet\/},  or 
simply, {\bfem $r$-multijet\/}, 
${}_kj^rf$ of $f:M\to\R$ at $(x_1,...,x_k)\in M^{(k)}$ is
\begin{align}
  {}_kj^rf(x_1,...,x_k):=\paren{j^rf(x_1),...,j^rf(x_k)}. 
\end{align}
The {\bfem $r$-multijet space} 
${}_kJ^r(M,N)$ is given by
\begin{align}
{}_kJ^r(M,N)&:=\{{}_kj^rf(x_1,...,x_k)\in
(J^r(M,N))^k|\quad (x_1,...,x_k)\in M^{(k)}\}. 
\end{align}
The map
$
  {}_kj^rf:M^{(k)}\to {}_kJ^r(M,N)
$
is called an {\em $r$-multijet section\/}. 
\label{def:mjet}
\end{definition}


\begin{definition}[Natural stratification of ${}_kJ^r(M,{\R})$]
The {\bfem natural stratification\/} of ${}_kJ^r(M, {\R})$ is the one
given by 
\begin{align}
{\cal S}({}_kJ^r(M,{\R})):=&\{\Delta_k\cap X_1\times ...\times X_k
|
X_1,...,X_k \in {\cal S}(J^r(M,{\R}))\}, 
\end{align}
where 
\begin{align}
\Delta_k := &\bigl\{ (j^rf_1(p_1),...,j^rf_k(p_k)) 
\in
{}_kJ^r(M,{\R}) \big|
f_1(p_1) = ... =f_k(p_k)\bigr\}. 
\end{align}
\end{definition}

Hence we find that the stable multiunfolding germ is given by multituple of well known
single stable unfolding under the aid of Mather's multitransversality theorem\cite{MA}. 
This does not change in the case with discrete symmetry since the multiunfolding is point-wise. 
With continuous symmetry we cannot use the multitransversality theorem directly since
a family of uncountable unfoldings is considered.


Finally we give a minimum function to investigate the concrete Maxwell set.
If the minimum function $\mu_F$ has a singularity at 
the point $u_0$, then the function $f_{u_0}$ has
several global minimal points on the manifold $M$.
This will give just the Maxwell set even if the minimum point is not isolated in the case of continuous symmetry.

\begin{definition}[Minimum function]
The {\em minimum function\/} ${\mf}:U\to\R$ of 
an unfolding $F: M\times U\to {\R}$ 
is given by 
\begin{equation}
  {\mf}(u):=\min \{F(x,u)| x\in M\}.
\end{equation}
\label{def:mmf}
\end{definition}

There are two cases when the minimum function ${\mf}$ has a singularity
at $u$:\\
(1) the function $f_{u}$ has
several  {\minimal}  points in $M$, or \\
(2) the number of minimum points changes there. \\
In the case (1) $u$ is a point of  the Maxwell set. 
In the case (2) $u$ is not a point of the Maxwell set but 
corresponds to a point of the endpoint set $\EPS$. 


\section{classification of the symmetric Maxwell set}

In the paper\cite{SK}, the stable Maxwell set of multiunfolding $F:(M\times U,M\times\{u_0\})\to {\R}$, 
where $M$ is two dimensional
manifold and $U$ is three-dimensional open subset, were mathematically classified.
There, based on well established local investigation of the unfolding\cite{PS}, semi-local classification
of the universal multiunfolding is made and it gives the classification of the Maxwell set. 
Then using multitransversality theorem, it is shown that they are stable.

On the other hand, some revisions are required for the present study in symmetric spacetime.
With any symmetry, we should impose a condition for the symmetry to the list of the standard functions which 
are representatives
of classification. Then diffeomorphism equivalence of the classification would be restricted to SPD equivalence 
under the request of the symmetry.
We realize the restriction by using some global coordinates related to the symmetry instead of the local coordinate
of neighborhoods.

That simply results from imposing a symmetry to the Maxwell set but revisions are not only this.
With any symmetry, we should change the concept of the stability of
the unfoldings from general ones to symmetric ones (Def.\ref{def:gsms}, Def.\ref{def:gsmp}) 
since the definitions of perturbation and the stable multiunfolding are altered by the SPD and the symmetric subset $G$-$\Cinf(M\times U,{\R})$.
Consequently, it is expected that a new class of multiunfolding will be added to the classification as a stable one.
This seriously embarrasses the problem.
Especially for the case of continuous symmetry, the function space where unfolding and perturbation
are defined changes its dimensions (strictly speaking, the dimensions of its jet space) by infinite dimensions. 
This implies that an excluded function germ
because of its infinite codimensions (which means the function germ cannot have universal unfolding) in function space $\Cinf(M\times U,{\R})$ is 
possible to be with finite codimensions and provide a stable unfolding in $G$-$\Cinf(M\times U,{\R})$. 
To find this directly, we should take a systematic survey of all functions in $G$-$\Cinf(M\times U,{\R})\sim \Cinf(\frac{M\times U}{G},{\R})$, and is not easy because $U/G$ has a boundary at the fixed points of $G$.
In addition to it, because of such continuous symmetry, the concept of the usual point-wise germ will
fail since the function germ orbits along the isometry. 
Especially at the fixed point of the isometry in $U$, their minimum points become non-isolated.

In the previous section, a $G$-orbital multigerm have been defined to resolve the latter problem (Def.\ref{def:gom}).
To avoid the former trouble, we first consider a discrete symmetry $G_d$ in low-dimensional spacetime $M', U'$ whose
catastrophe is essentially equivalent to the continuous symmetry $G_c$ because of $U/G_c\sim U'/G_d$.
For example, the reflection symmetry in $(n-1)$-dimensional spacetime and
axial symmetry in $n$-dimensional spacetime are so.
Since the change from the general function space $\Cinf(M'\times U',\R)$ to the discrete symmetric 
function space $G_d$-$\Cinf(M'\times U',\R)$ is exceedingly small so that one can
systematically pick up multiunfoldings newly added to the stable classes.

First, we recall a symmetry condition about the unfolding.
We suppose the spacetime $(\MM,g)$ is symmetric and there exists a group of isometry $G\ni\varphi$,
 $\varphi^*g=g$ (eq.(\ref{eqn:isom})).
As stated in the section 2, $M$ and $U$ are invariant subspace by the construction,
\begin{equation}
\varphi(M)= M
,\ \ \ \varphi(U)= U,\ \ \ \varphi(M\times U)= M\times U.
\end{equation}

As the discussion in the section 2, 
the unfolding is pulled back by $\varphi$,
and should be invariant under the isometric transformation,

\begin{equation}
F(x,u)=\varphi^*(F)(x,u)=F(\varphi(x),\varphi(u)).
\label{eqn:smuf}
\end{equation}
Therefore, we confirm that the Maxwell set should be invariant by the isometry group.
To find a classification of the Maxwell set in this symmetric spacetime, we will list the $G$-stable multiunfolding.
It should be a $G$-orbital multigerm of unfolding (Def.\ref{def:gom}).

\subsection{discrete symmetry}

In the case of discrete symmetry investigation is easy to perform.
Here, we suppose the isometry group is a discrete group $G_d\ni \varphi_d$.

\subsubsection{Class A multiunfolding}

If the symmetry is discrete, as long as there is no fixed point, the classification of stable
multiunfolding is essentially same as the case without symmetry\cite{SK}. Without fixed points, locally
$\frac{\UU_M\times\UU_U}{G}\sim \frac{\UU_M}{G}\times \UU_U$, where $\UU_M\subset M$ and $\UU_U\subset U$
is a open subset of $M$ and $U$, respectively.
Then the identification
\begin{equation}
{}_kJ^r\left(\frac{\UU_M\times\UU_U}{G},\R\right)\sim {}_kJ^r\left(\frac{\UU_M}{G}\times \UU_U,\R\right),
\label{eqn:jse}
\end{equation}
implies that the catastrophe of this case is equivalent to the catastrophe without symmetry.

By the action of the isometry (\ref{eqn:smuf}) on an unfolding $F$, we generate a $G_d$-invariant $G_d$-orbital 
multigerm of unfolding
from a multigerm of unfolding since $G_d$-orbital multigerm is point-wise germ. We write this $G_d$-invariant 
$G_d$-orbital multigerm of unfolding as $G_d[F]$.

\begin{align}
G_d[F]&=[G_d[F]]_{\{((x_1,...,x_k),u_0),((\phi_d(x_1),...,\phi_d(x_k)),\phi_d(u_0)),...|\phi_d\in G_d\}}\\
&=[[F]_{((x_1,...,x_k),u_0)},[\phi_d^*F]_{(\phi_d(x_1),...,\phi_d(x_k)),\phi_d(u_0)},...]=[F,\phi_d^*F,...], \\
\phi_d&\in G_d.
\label{eqn:gf}
\end{align}
In $[X_1,X_2,...]$, $[\ ]$ means that each element $X_i$ does not share $u_0$ and is not competent to others,
differently from $(X_1,X_2,...)$.

There is no need to care the change of the function space (\ref{eqn:jse}).
To get a representative function of the classification we simply use a function $F$ in the case
of the classification without symmetry.

Though $u_0$ is mapped to $\varphi_d(u_0)$ in (\ref{eqn:gf}), there is no more degrees of freedom for
 control parameter space $U$ of $G_d[F]$ than those of
the original multiunfolding $F$.
Of course it is obviously $G_d$-stable at the minimum points if the original $F$ is stable at the minimum points. 
The reason is because perturbation is considered also in ${}_kJ^r(\frac{\UU_M\times \UU_U}{G},\R)$.
We call this a Class A multiunfolding in the following.

\subsubsection{Class B multiunfolding}

Next we consider the case with fixed points of the isometry on $M$ or $U$ in the neighborhood of minimum points $G_d((p_1,p_2,...p_k),u_0)$.
Though this point in $M^{(k)}$ or  $U$ is the fixed point of only several elements $g_f\in G$ in general,
considering the subgroup $\langle g_f\rangle=G_f\subseteq G$ generated by them, we suppose the fixed point is of
all elements of the isometry in the present article. 
General case would be developed by combining the discussion of both cases with and without fixed points.

If a minimum point $p_i$ for $u_0$ is a fixed point of $\varphi_d$,
$u_0$ should also be on a fixed point of $\varphi_d$ in our discussion.
This is required in the case of an event horizon.
A couple of $u_0$ and a minimum point $p_i(u_0)$ determines a generator $(u_0,p_i)$ of the event horizon. 
Nevertheless, from the symmetry also $(\varphi(u_0),\varphi(p_i)=p_i)$ is a generator of the same event horizon.
This means two generators pass a one point $p_i$ on $\CS_{St_1}\cap\HH=\psi(M)$. Nevertheless this is not 
allowed since there is no future endpoint on the event horizon.
Then we suppose $u_0$ is always a fixed point of the isometry here.

In this case, we must pay attention to the changes of the function space from $\Cinf(M\times U,\R)$ to $G_d$-$\Cinf(M\times U,\R)$, where unfolding and perturbation
is given since the essential control space $U/G_d$ has boundary at the fixed point.
To care this problem, it is convenient also to represent a
standard function by not local coordinates but global coordinates to indicate the symmetry.
By the action of the isometry to the global coordinate, we observe that the codimension of the stratum for multijet space (Def.\ref{def:mjet}) is
possible to be diminished since $u_0$ is common among multituple of multiunfolding for the $G_d$-orbital multigerm of unfolding.
When we list the strata for the multijet space, in advance, we should take extra strata assuming the decreasing of the control parameter.
By this, there is possibility that a new multiunfolding becomes stable.

We find $G_d$-stable multiunfoldings by the following steps.

\begin{enumerate}
\item  One prepares strata for a single-jet space which possess minimum points from Def.\ref{def:str} and
give its universal unfolding, so that the number of control parameter does not exceed the dimension of $U$. 
This is well understood and is known to be stable by Thom's theorem in the context of elementary catastrophe.
They are written in local coordinates.
\item List the possible multituples of unfoldings and sum the numbers of their control parameters.
 By the isometric transformation of unfolding (\ref{eqn:smuf}), some of control parameters are identified in the global
coordinate for the symmetry. 
Then we manage the list so that also the sum of the number of control parameters does not exceed the dimension of $U$
 after such identifications.
\item Imposing the symmetry relation (\ref{eqn:smuf}) on a possible globalizations of coordinate,
we will find a $G_d$-invariant $G_d$-orbital multigerm of unfolding.
Of course any of the multituples may not generate $G_d$-invariant $G_d$-orbital multigerm of unfolding.
\item 
In the same logic as in \cite{SK}, it can be shown that the $G_d$-invariant $G_d$-orbital multigerm of unfolding
 constructed above, 
is the $G_d$-stable at the minimum points, since each unfolding at a minimum points is $G_d$-stable unfolding even after imposing the symmetry. 
This is guaranteed by the fact the jet section $j^rF:M\times U\mapsto J^r(M\times U,\R)$ transversally crosses the strata also in $G_d$-$\Cinf(M\times U,\R)$ as long as the projection $\rho:\Cinf(M\times U,\R)\mapsto G_d$-$\Cinf(M\times U,\R)$ maps neither the stratum nor the jet section degenerate into a point.
It is easily checked that the condition (\ref{eqn:smuf}) never does so in most of significant cases. 
\end{enumerate}

Since the $G_d$-orbital multigerm of unfolding found in this way is $G_d$-invariant, it should be provided as the one generated
by the isometry group $G_d$. Nevertheless it includes competence at the fixed point
$u_0$ among multiunfoldings generated by each elements $\varphi_d\in G_d$, and is not $G_d[F]$.
Then we write such a $G_d$-invariant $G_d$-orbital multigerm of unfolding as $G_d(F)$;
\begin{align}
G_d(F)&=[G_d(F)]_{\{((x_1,...,x_k),u_0),((\phi_d(x_1),...,\phi_d(x_k)),u_0),...|\phi_d\in G_d\}}\\
&=([F]_{((x_1,...,x_k),u_0)},[\phi_d^*F]_{(\phi_d(x_1),...,\phi_d(x_k)),u_0},...)=(F,\phi_d^*F,...) ,\\
 \phi_d&\in G_d.
\end{align}
$F,\phi_d^*F,...$ share $u_0$ and compete to each other.
We call this a Class B multiunfolding.

\subsection{continuous symmetry}
Now we write continuous isometry group as $\varphi_c\in G_c$.
To avoid some difficulties caused by the continuous symmetry, here we suppose that the catastrophe of the case of this
continuous symmetry is equivalent to a catastrophe with any discrete symmetry. 
Here an essential control parameter space is not $U$ rather $U/G$ since the isometry does not allow the unfolding in the 
direction of the isometry.
Then essentially the classification of stable multiunfolding with the continuous symmetry is already completed,
if such a corresponding discrete symmetry really exists and if their catastrophe is equivalent, since in the previous subsection, we have investigated the procedure to classify the $G_d$-stable $G_d$-orbital multigerm of 
unfolding.

In the next section, it will be shown that there is identification
between the control space for a reflection symmetry in three-dimensional spacetime
and that for an axial symmetry in four-dimensional spacetime.
Since their catastrophes are equivalent, we only have to translate the classification of the multiunfolding with discrete symmetry
into that with continuous symmetry.

\subsubsection{Class C multiunfolding}

To find a $G_c$-invariant $G_c$-orbital multigerm of unfolding for the continuous symmetry, it is an easy way
to generate it by the isometry from the unfolding of discrete symmetry as a orbit of the continuous isometry.
When there is no fixed point in the neighborhood of $((p_1,...,p_k),u_0)$, it is possible to generate $G_c$-invariant 
$G_c$-orbital multigerm of unfolding from a $G_d$-invariant $G_d$-orbital multigerm of unfolding by straightforward 
correspondence.

Here one should be careful to start from a multiunfolding in Class A.
To find a $G_d$-invariant $G_d$-orbital multigerm of unfolding in Class A, it was generated by the isometry group as
 $G_d[F]$.
Nevertheless, in the continuous symmetry, $G_c$-orbital multigerm suffers overlap of neighborhoods $\UU_1$ and $\UU_2$ 
generated by an isometry $\varphi_c(\UU_1)=\UU_2$. Therefore the multigerm $F(\varphi_c(x),\varphi_c(u))$ may be 
inconsistent to $F(x,u)$ on $\UU_1\cap\UU_2$.
To resolve this inconsistency it is sufficient that the original unfolding $F$ is invariant along the continuous isometry in $\UU_1$. This also is realized by the global coordinate of the continuous symmetry.

To represent the equivalence class of $G_d$-orbital multigerm, multituple notation was possible in the discrete symmetry.
On the other hand, in this case of continuous symmetry, we write it as a family of uncountable members,

\begin{align}
G_c[F]&=[G_c[F]]_{\{(\varphi_a(x_1),...,\varphi_c(x_k)),\varphi_c(u_0)|\varphi_c\in G_c\}} \\
&=[\{[\varphi_c^*F]_{(\varphi_a(x_1),...,\varphi_c(x_k)),\varphi_c(u_0)}|\varphi_c\in G_c\}]=[\{\varphi_c^*F|\varphi_c\in G_c\}].
\end{align}
Since original $F$ is invariant under the isometry $\varphi_c^*F=F$ in global coordinate it is sufficient to write as $[F|G_c]$.
 
Now we need to revive a dummy variable $u_t$ of $U$, which is omitted by the quotient $U/G_c$, to decide the Maxwell set in $U$. 
Since the generator of isometry $\partial /\partial t$  should not effect the catastrophe, that can be achieved by adding a Morse function from the splitting lemma\cite{PS}.
Therefore it is sufficient to add a $\delta t^2$ with $\varphi_c(\delta t)=\delta t$ in order to find $G_c$-invariant $F$. 
The dummy variable $u_t$ is included in this as $\delta t=\delta t(u_t)$.
This case is named as a Class C multiunfolding.

\subsubsection{Class D multiunfolding}

At the fixed point, also the $G_c$-invariant $G_c$-orbital multigerm of unfolding is generated by the continuous isometry since the fixed point of $G_c$ is same as that of $G_d$.
Using global coordinates, it is possible to rewrite such multiunfoldings in symmetric form,

\begin{align}
G_c(F)&=[G_c(F)]_{\{(\varphi_a(x_1),...,\varphi_c(x_k)),u_0)|\varphi_c\in G_c\}} \\
&=(\{[\varphi_c^*F]_{(\varphi_a(x_1),...,\varphi_c(x_k)),u_0)}|\varphi_c\in G_c\})=(\{\varphi_c^*F|\varphi_c\in G_c\}).
\end{align}
Since original $F$ is $G_c$-invariant just like the Class C multiunfolding, it is sufficient to write $(F)$.
Here $(F)$ should be written as a standard function using a global coordinate in the direction of the continuous isometry
to clarify that its minimum points make an orbit along the isometry. This will be found by discussion in a following
 concrete situation.
Instead of the Morse function, a global function which is locally right SPD equivalent to a Morse function are added to 
revive a dummy variable (see next section).

By the splitting lemma\cite{PS}, there is no need to discuss the transversality and stability, since it is supposed that
the essential control parameter space $U/G_c$ for the continuous symmetry 
is identical to that of the discrete symmetry $U/G_d$ in low-dimensional spacetime where the stability have been studied.

These are class D multiunfoldings.

\section{concrete classification: reflection symmetry and axial symmetry}
\subsection{reflection symmetry in three-dimensional spacetime}

The main purpose of the present article is to classify the stable crease set of an event horizon in an axially
symmetric spacetime.
To do so, we study Maxwell sets in reflection symmetric three-dimensional spacetime where $U$ is two dimensional 
and $M$ is one-dimensional, in advance.

A reflection isometry form a ${\bf Z}_2$ group, that is
\begin{align}
\varphi_r&:M\mapsto M, U\mapsto U,\\
 G_r&=\{id, \varphi_r|\varphi_r\circ \varphi_r=id\}\simeq {\bf Z}_2,
\end{align}
where $id$ is the identical map.

\subsubsection{Class A multiunfolding with reflection symmetry}

First, we consider a reflection symmetric $G_r$-orbital multigerm of unfolding $F_{G_r}$ without a fixed point (a Class A multiunfolding).
As discussed in the previous section, a $G_r$-invariant $G_r$-orbital multigerm of unfolding $G_r[F]$ is generated by the
reflection isometry $\varphi_r$ from a usual multiunfolding $F$ and is $G_r$-stable at the minimum points if the original
multiunfolding $F$ is stable at the minimum points.

We put the classification of multiunfolding without symmetry, that is the case of $codim\leq 4$ from \cite{SK}\footnote{
Since the dimensions of $M$ is different, the $codim$ is differ by one between this case and \cite{SK}}, so that they
are also stable in the case with two-dimensional $U$.
A stable multiunfolding $F$ is right equivalent to one of the following multiunfolding germ at their minimum points.
\begin{itemize}
\item $A_3=x^4+u_2x^2+u_1x, \ \text{at } (0,0)\in\R\times\R^2$.
\item $(A_1,A_1)=(x_1^2+u_1,x_2^2), \ \text{at } ((0_1,0_2),0)\in \R^2\times\R^2$ where $0_i$ is the origin of local coordinate $(x_i,y_i)$.
\item $(A_1,A_1,A_1)=(x_1^2+u_1,x_2^2+u_2,x_3^2),
 \text{at } ((0_1,0_2,0_3),0)\in\R^3\times\R^2$.
\end{itemize}
where $A_i$ means a stable unfolding of the stratum $A_i$.
The unfolding $A_1$ is omitted since its Maxwell set is empty set.
Then, we give reflection symmetric $G_r$-orbital multigerms of unfolding as $G_r(F)$.

{\bf Class A multiunfolding}

A $G_r$-invariant $G_r$-stable multiunfolding in Class A is right SPD-equivalent to one of the following $G_r$-orbital multigerms of unfolding at
the minimum points.
\begin{itemize}
\item $G_r[A_3]=[A_3,\varphi_r^*A_3]=[A_3(x,u),A_3(\varphi_r(x), \varphi_r(u))]$.
\item $G_r[(A_1,A_1)]=[(A_1,A_1),\varphi_r^*(A_1,A_1)]=[(A_1,A_1),(A_1(\varphi_r(x), \varphi_r(u)),A_1(\varphi_r(x), \varphi_r(u)))]$. 
\item $G_r[(A_1,A_1,A_1)]$$=[(A_1,A_1,A_1),\varphi_r^*(A_1,A_1,A_1)]$\\    $=[(A_1,A_1,A_1),(A_1(\varphi_r(x), \varphi_r(u)),A_1(\varphi_r(x), \varphi_r(u)),A_1(\varphi_r(x), \varphi_r(u)))]$.
\end{itemize}

\begin{figure}
\begin{center}
\includegraphics[width=11cm,clip]{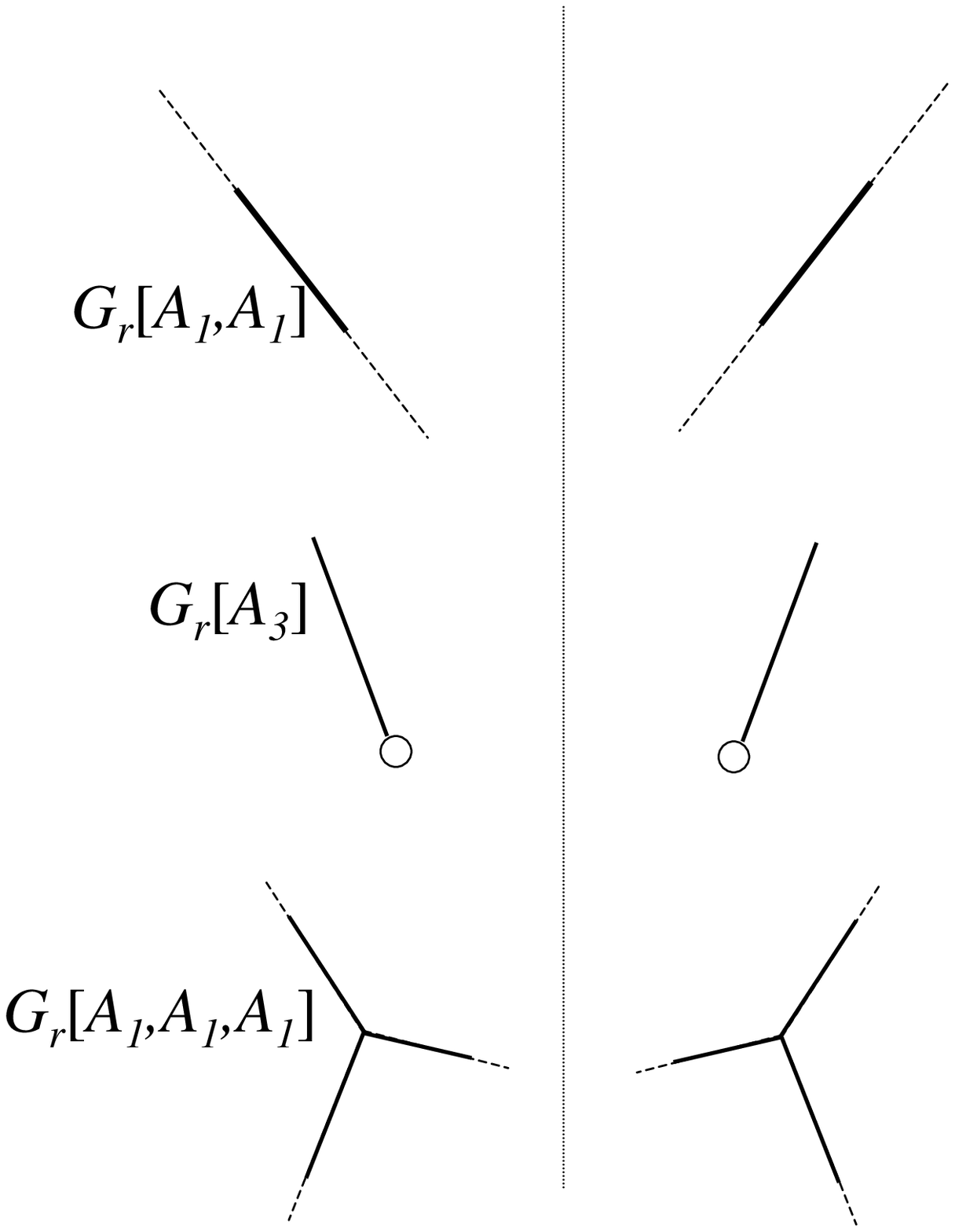} 
\end{center}
\caption{The Maxwell sets in two-dimensional reflection symmetric spacetime not on its fixed points. 
They are the Maxwell sets of Class A multiunfoldings.}
\label{fig:mcA}
\end{figure}

Since the function space of perturbation is same as that of without symmetry in the neighborhood of $((p_1,...,p_k),u_0)$,
these are all of the stable multiunfoldings in this class.
Their Maxwell sets are given by the Maxwell sets in \cite{SK} and their reflectional images.
They are illustrated in Fig.\ref{fig:mcA}.

\subsubsection{Class B multiunfolding with reflection symmetry}

Next we consider  Class B multiunfoldings, where $u_0$ is on a fixed point of $\varphi_r$.
For convenience, we introduce global coordinates $(x,u_x,u_z)\in M\times U$ by which the reflection symmetry can be represented. 
The reflection isometry is given by  $(x,u_x,u_z)\mapsto (-x,-u_x,u_z)$ in our convention
of the coordinates.

We will find the list of the stable multiunfoldings in a procedure developed in the previous section.
\begin{enumerate}
\item Since we treat only global minimums and $M$ is one- and $U$ is two-dimensional,
only $A_1$ and $A_3$ are considered among the list of the strata (Def.\ref{def:str}), whose codimension is less than 
four and the number of control
parameter is not more than two ($=dim(U)$). Their stable unfoldings are $A_1=x^2$, and $A_3=x^4+u_2x^2+u_1x$.
\item Since even after one imposes the reflection symmetry the codimension of the strata for multijet space 
$\Delta_k\cap(A_i\times ...)$ does not decrease
more than $[codim/2]$ ($[x]$ is the Gauss's symbol), it is sufficient
to consider the case in which codimension of $\Delta_k\cap(A_i\times ...)$ is less than five 
(the number of control parameters
is less than four), where the minimum value is fixed by the reflection symmetry.
The list of the $k$-tuple of the stable unfoldings is
\begin{eqnarray}
(A_1,A_1)&=&(x_1^2+u_1,x_2^2)\\
(A_1,A_1,A_1)&=&(x_1^2+u_1,x_2^2+u_2,x_3^2)\\
(A_1,A_1,A_1,A_1)&=&(x_1^2+u_1,x_2^2+u_2,x_3^2+u_3,x_4^2)\\
(A_3)&=&(x_1^4+u_2x_1^2+u_1x_1)\\
(A_3,A_1)&=&(x_1^4+u_2x_1^2+u_1x_1,x_2^2+u_3).
\end{eqnarray}
\item Imposing the symmetry relation (\ref{eqn:smuf}) by the global coordinate, 
the above $k$-tuples of unfoldings become the following $G_r$-invariant $G_r$-orbital multigerms of unfolding 
if the number of
their control parameters $u_i$
finally is not exceeding $\dim(U)$. $(A_3,A_1)$ is not allowed since the number of parameter is three even after imposing symmetry.
\item We can concretely confirm that projection $\rho:\Cinf(M\times U,\R)\mapsto G_r$-$\Cinf(M\times U,\R)$ 
maps neither the strata nor jet section into a point.
Therefore the following $G_r$-invariant $G_r$-orbital multigerms of unfolding $G_r(F)$ are $G_r$-stable at the minimum points,
since their multijet section transversally crosses the strata $\Delta_k\cap(A_i\times ...)$ in the jet space of $G_r$-$\Cinf(M\times U,\R)$.
This will be proved in the same logic as in \cite{SK}. 
\end{enumerate}

{\bf Class B multiunfolding}

A $G_r$-invariant $G_r$-stable multiunfolding in Class B is right SPD-equivalent to one of the following $G_r$-orbital multigerms of unfolding at
the minimum points.
\begin{itemize}
\item $(A_1,A_1)=(x_{r1}^2+u_z,x_{r2}^2)$.

To find the Maxwell set of it, we determine the minimum function (Def.\ref{def:mmf}) of 
the $G_r$-orbital multigerm of unfolding,
\begin{equation}
\min(A_1,A_1)=\begin{cases}
u_z &  u_z<0\\
0 & u_z\geq 0
\end{cases}.
\end{equation}
Then the Maxwell set that is the singular point of the minimum function
is 
\begin{equation}
\Mset(A_1,A_1)=\{(u_x,u_z)|u_z=0\}.
\end{equation}

\item $G_r(A_1)=(A_1,\varphi_r^*A_1)=((x_r-x_0)^2+u_x,(x_r+x_0)^2-u_x)\sim (x_1^2+u_x,x_2^2-u_x)$,\\
 where $x_1$ and $x_2$ are local coordinates.

The minimum function is
\begin{equation}
\min(A_1,\varphi_r^*A_1)=\begin{cases}
-u_x &  u_x<0\\
u_x & u_x\geq 0
\end{cases}.
\end{equation}

The Maxwell set that is the singular point of the minimum function
is
\begin{equation}
\Mset(A_1,\varphi_r^*A_1)=\{(u_x,u_z)|u_x=0\}.
\end{equation}

\item $(G_r(A_1),A_1)=((A_1,\varphi_r^*A_1),A_1)=(x_1^2+u_x+u_z,x_2^2-u_x+u_z,x_{r3}^2)$.

The minimum function is
\begin{equation}
\min((A_1,\varphi_r^*A_1),A_1)=min(-|u_x|+u_z,0)=\begin{cases}
|u_x| &  |u_x|>u_z\\
0 & |u_x|\leq u_z
\end{cases}
\end{equation}

The Maxwell set that is the singular point of the minimum function
is
\begin{align}
\Mset((A_1,\varphi_r^*A_1),A_1)&=\{(u_x,u_z)|u_x=0,u_z<0\}\nn
&\cup\{(u_x,u_z)||u_x|=u_z\}.
\end{align}

\item $G_r((A_1,A_1))=(G_r(A_1),G_r(A_1))=((A_1,\varphi_r^*A_1),(A_1,\varphi_r^*A_1))$\\
$=((x_1^2+u_x,x_2^2-u_x),(x_3^2+u_x+u_z,x_4^2-u_x+u_z))$,\\
where $x_1,x_2,x_3,x_4$ are local coordinates.

The minimum function is
\begin{equation}
\min((A_1,\varphi_r^*A_1),(A_1,\varphi_r^*A_1))=min(-|u_x|,u_z-|u_x|)=
\begin{cases}
-|u_x|+u_z &  u_z<0\\
-|u_x| & |u_z\geq 0
\end{cases}.
\end{equation}

The Maxwell set that is the singular point of minimum function
is
\begin{align}
\Mset((A_1,\varphi_r^*A_1),(A_1,\varphi_r^*A_1))&=\{(u_x,u_z)|u_x=0\}\nn
&\cup\{(u_x,u_z)|u_z=0\}.
\end{align}

\item $G_r(A_3)=A_3=x_r^4+u_zx_r^2+u_xx_r$.

The Maxwell set is same as that of $A_3$ without symmetry.
\begin{equation}
\Mset(A_3)=\{(u_x,u_z)|(u_z< 0,u_x=0)\}.
\end{equation}
\end{itemize}

\begin{figure}
\begin{center}
\includegraphics[width=11cm,clip]{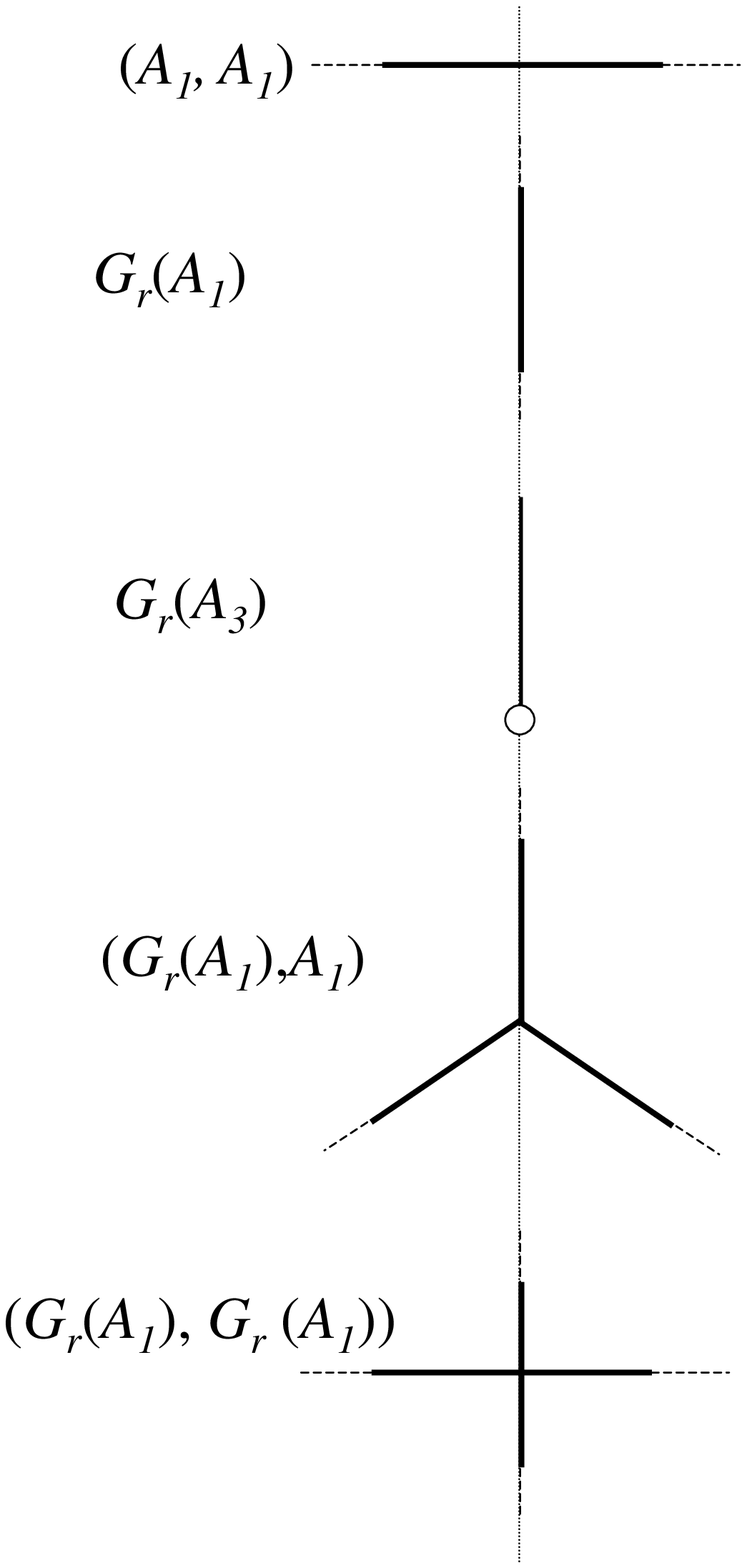} 
\end{center}
\caption{The Maxwell sets in two-dimensional reflection symmetric spacetime on its fixed points. 
They are the Maxwell sets of Class B multiunfoldings.}
\label{fig:mcB}
\end{figure}

It should be noted that some multiunfoldings decreases the number of control parameter after imposing the 
reflection symmetry.
Especially, it is note worthy that $((A_1,\varphi_r^*A_1),(A_1,\varphi_r^*A_1))$ is not stable without the reflection symmetry but stable
 with the reflection symmetry. Then the entirely new Maxwell set have been added to the classification.
These Maxwell sets are illustrated in Fig.\ref{fig:mcB}

\subsection{axial symmetry in four-dimensional spacetime}
Here we consider axial symmetry in a four dimensional spacetime.
In this case, its isometry group $G_a$ is continuous group and isomorphic to $U(1)$,
\begin{align}
\varphi_{\theta}&:M\mapsto M, U\mapsto U, \varphi_{\theta_0}=exp\left[\theta_0\left(\frac{\partial}{\partial\phi}\right)\right]\nn
 G_a&=\{\varphi_{\theta}|0\leq\theta\leq 2\pi, \varphi_{2\pi}=id\}\simeq U(1),
\end{align}

\subsubsection{Class C multiunfolding with axial symmetry}

For the reflection symmetry, the function space was restricted by only finite codimensions
and there does not appear an essentially new unfolding without a fixed point.
As the axial symmetry, however, is continuous, our standard investigation will fail by two reasons.
Since the function space (strictly speaking, its jet space) is restricted into infinite codimensional
subspace,
a new function which never provides stable unfolding owing to its infinite codimension
 can provide a stable unfolding, e.g. $(x^2+y^2)^2$ see \cite{MS2}.
To find all of these new functions it is required to survey the whole right SPD equivalence class of the restricted 
function space 
systematically and
it is embarrassing.

Another is that infinite number of
stationary points can be generated by the axial isometry as $\varphi_{a}^*F$. 
Then we cannot use semi-local discussion where usual function germ
defined.
Nevertheless, the essential control space  of this axial symmetry  in four dimensional spacetime $U/G_a$ is identical to 
that of reflection symmetry in three-dimensional spacetime $U'/G_r$.
Omitting the angular direction of the state space $M$ and control space $U$, we treat the remaining
as a state
space and control space $U/G_a$. We can study it in semi-local formalism.

To determine a correspondence between $U_r=(U'\sim\R^2)/G_r$ and $U_a=(U\sim\R^3)/G_a$, it is convenient to use
coordinate systems $(u_x',u_z')$, $(u_r,u_{\theta},u_z)$.
By identifications $u_x'=u_r$ and $u_z'=u_z$, a correspondence $\R^2/G_r\mapsto \R^3/G_a$ is given,
since the conditions $0\leq u_x'<\infty, -\infty<u_z'<\infty$ are consistent to 
$0\leq u_r<\infty, -\infty<u_z<\infty$.
This correspondence simply gives the correspondence between Class A multiunfolding and the Class C multiunfolding.
We also use local coordinate $(x_i,u_i)$  to show local diffeomorphism equivalence. 
$x_i$ and $u_i$ are local coordinate in $r$-$z$ or $u_r$-$u_z$ section of $M$ and $U$, respectively.

To determine a Maxwell set we should give a unfolding in $U$ not in $U/G_a$.
From the splitting lemma\cite{PS}, we understand that it is sufficient to add a Morse function by a dummy variable
 $u_{\theta}$ of angular direction,
 which does not concern the structure
of the minimum point,
as $\delta\theta^2=(\theta-u_\theta)^2$ where $\theta-u_\theta$ is considered to be small\footnote{ One should not translate this
as the generator of an event horizon does not change its angular direction. To write this form, we may change the origin or scale of the angular 
coordinate $\theta$}.

{\bf Class C multiunfolding}

A $G_a$-invariant $G_a$-stable multiunfolding in Class C is right SPD-equivalent to one of the following $G_a$-orbital multigerm of the unfolding at
the minimum points.
\begin{itemize}
\item $G_a[(A_1,A_1)] =(x_1^2+y_1^2+u_1+(\theta_1-u_{\theta})^2,x_2^2+y_2^2+(\theta_2-u_{\theta})^2)$, \\
$0\leq\theta_1\leq 2\pi$, $0\leq\theta_2\leq 2\pi$, $0\leq u_{\theta}\leq 2\pi$.

For $G_a[(A_1,A_1)]$, the minimum function is
\begin{equation}
\min(G_a[(A_1,A_1)])=\begin{cases}
u_1 &  u_1<0\\
0 & u_1\geq 0
\end{cases}.
\end{equation}
Then its Maxwell set is given by
\begin{equation}
\Mset(G_a[(A_1,A_1)])=\{(u_1,u_2,u_{\theta})|u_1=0\}.
\end{equation}

\item
$G_a[(A_1,A_1,A_1)]=(x_1^2+y_1^2+u_1+(\theta_1-u_{\theta})^2,x_2^2+y_2^2+u_2+(\theta_2-u_{\theta})^2,x_3^2+y_3^2+(\theta_3-u_{\theta})^2)$, \\
$0\leq\theta_1\leq 2\pi$, $0\leq\theta_2\leq 2\pi$, $0\leq\theta_3\leq 2\pi$, $0\leq u_{\theta}\leq 2\pi$.

For $G_a[(A_1,A_1,A_1)]$, the minimum function is
\begin{equation}
\min(G_a[(A_1,A_1,A_1)])=\begin{cases}
u_1 &  u_1<u_2,u_1<0\\
u_2 & u_2<u_1,u_2<0 \\
0 & u_1>0,u_2>0 \\
\end{cases}.
\end{equation}
This is singular at $u_1=u_2<0$, $u_1=0, u_2>0$ and $u_2=0, u_1>0$.
Then its Maxwell set is given by 
\begin{align}
\Mset(G_a[(A_1,A_1,A_1)])&=\{(u_1,u_2,u_{\theta})|u_1=u_2<0\}\nn
&\cup \{(u_1,u_2,u_{\theta})|u_1=0, u_2>0\}\cup\{(u_1,u_2,u_{\theta})|u_2=0, u_1>0\}.
\end{align}

\item $G_a[A_3]= x_1^4+u_2x_1^2+u_1x_1+(\theta_1-u_{\theta})^2$, \\
$0\leq\theta_1\leq 2\pi$, $0\leq u_{\theta}\leq 2\pi$.

For $G_a[A_3]$, the Maxwell set is same as that of $A_3$ on the section $u_{\theta}=0$.
Then its Maxwell set is given by 
\begin{equation}
\Mset(G_a[A_3])=\{(u_1,u_2,u_{\theta})|u_2>0,u_1=0\}
\end{equation}
\end{itemize}

\begin{figure}
\begin{center}
\includegraphics[width=11cm,clip]{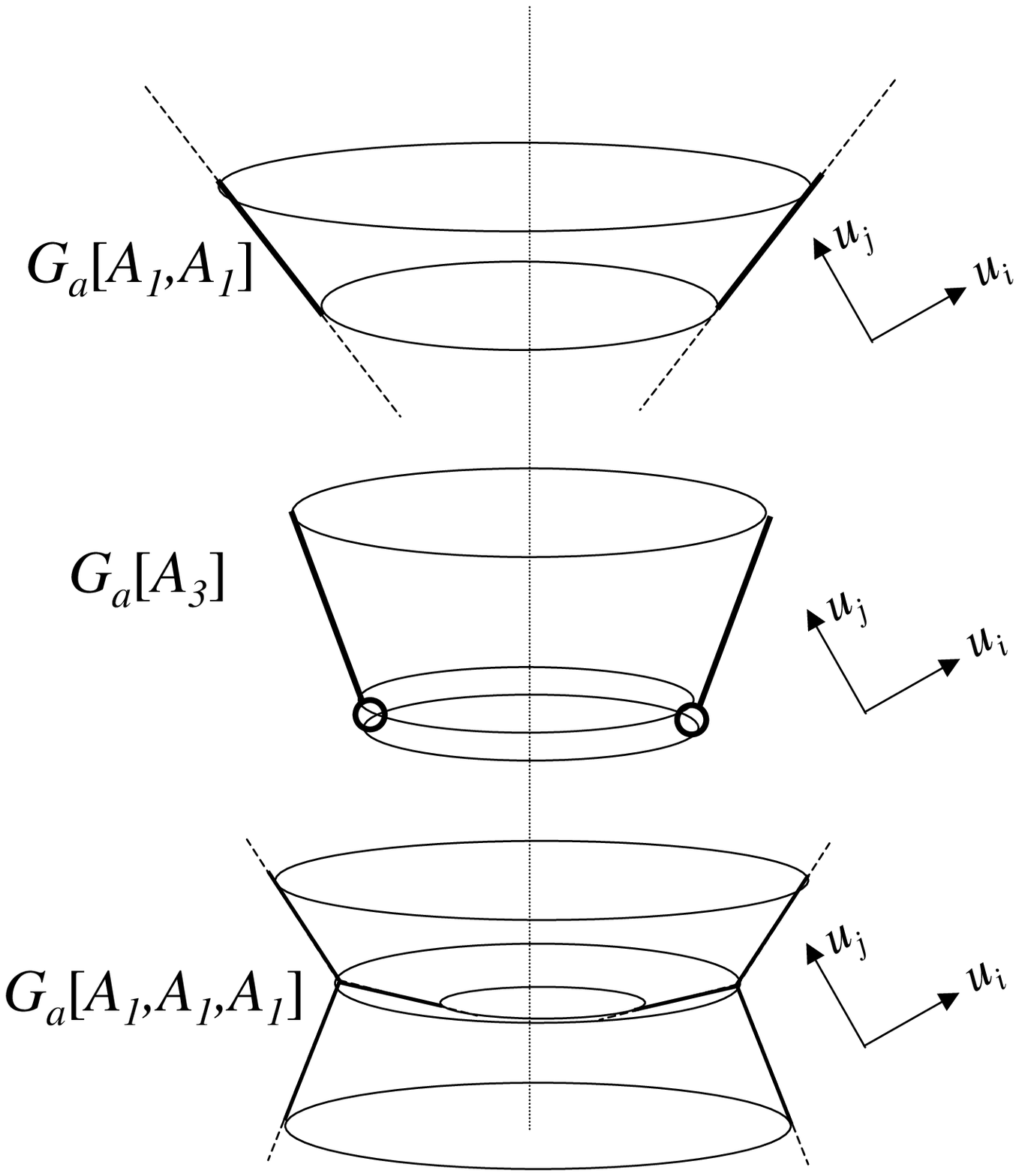} 
\end{center}
\caption{The Maxwell sets in three-dimensional axially symmetric spacetime not on its fixed points. 
They are the Maxwell sets of Class C multiunfoldings.}
\label{fig:axcA}
\end{figure}

These Maxwell sets are illustrated in Fig.\ref{fig:axcA}.

\subsubsection{Class D multiunfolding with axial symmetry}

Similarly to the Class C multiunfolding, the correspondence between $U_r=\R^2/G_r$ and $U_a={\R}^3/G_a$ also gives
correspondence of multiunfolding between Class B and Class D.
In addition that, it is necessary to show that the minimum is not isolated but make an orbit 
of the isometry.
This is realized by global coordinates of the symmetry so that the Morse function $\delta\theta^2$, which is added to revive
a dummy variable $u_{\theta}$, is 
extended to a global function of $\delta\theta$. It is achieved by $\cos\delta\theta\sim 1-\delta\theta^2/2+...$.

{\bf Class D multiunfolding}

A $G_a$-invariant $G_a$-stable multiunfolding in Class D is right SPD-equivalent to one of the following $G_a$-orbital multigerm of unfolding at
the minimum points.

\begin{itemize}
\item $(A_1,A_1)=(r_1^2+u_z,r_2^2)$.

For $(A_1,A_1)$, the minimum function is
\begin{equation}
\min(A_1,A_1)=\begin{cases}
u_z &  u_z<0\\
0 & u_z\geq 0
\end{cases}.
\end{equation}
The Maxwell set that is the singular point of minimum function
is 
\begin{equation}
\Mset(\varphi_{\theta}A_1)=\{(u_r,u_z,u_{\theta})|u_z=0\}.
\end{equation}

The toroidal event horizon reported in ref.\cite{ST} is caused by the crease set which is embedding
of this Maxwell set.

\item $G_a(A_1)= z^2+u_r\cos(\theta-u_{\theta})$, \\
$0\leq\theta_\leq 2\pi$,  $0\leq u_{\theta}\leq 2\pi$.

For $G_a(A_1)$, the minimum function is
\begin{equation}
\min(G_a(A_1))=u_r.
\end{equation}
This is singular at $u_r=0$,
\begin{equation}
\Mset(G_a(A_1))=\{(u_r,u_z,u_{\theta})|u_r=0\}.
\end{equation}

This Maxwell set is reported in the collision of the two black holes\cite{NCSA}.

\item $(G_a(A_1),A_1)=(z^2+u_r\cos(\theta-u_{\theta}),r_2^2+u_z)$,\\
 $0\leq\theta_\leq 2\pi$,  $0\leq u_{\theta}\leq 2\pi$

The minimum function of  $(G_a(A_1),A_1)$ is
\begin{equation}
\min(G_a(A_1),A_1)=min(u_r,u_z)=\begin{cases}
u_z &  u_r>u_z\\
u_r & u_r\leq u_z
\end{cases}.
\end{equation}
This is singular at $u_r=0, u_z>0$ and $u_z=u_r,u_z<0$;
\begin{align}
\Mset(G_a(A_1))&=\{(u_r,u_z,u_{\theta})|u_r=0, u_z>0\}\nn
&\cup \{(u_r,u_z,u_{\theta})|u_z<0,u_z=u_r\}.
\end{align}

\item $(G_a(A_1),G_a(A_1))=(z_1^2+u_r\cos(\theta_1-u_{\theta})+u_z,z^2+u_r\cos(\theta_2-u_{\theta}))$, \\
$0\leq\theta_1\leq 2\pi$, $0\leq\theta_2\leq 2\pi$, $0\leq u_{\theta}\leq 2\pi$.

The minimum function of $(G_a(A_1),G_a(A_1))$ is
\begin{equation}
\min((G_a(A_1),G_a(A_1)))=\min(u_r+u_z,u_r)=\begin{cases}
u_r &  u_z>0\\
u_r+u_z & u_z\leq u_z0
\end{cases}.
\end{equation}
This is singular at $u_r=0$ and $u_z=0$;
\begin{align}
\Mset((G_a(A_1),G_a(A_1)))&=\{(u_r,u_z,u_{\theta})|u_r=0\}\nn
&\cup \{(u_r,u_z,u_{\theta})|u_z=0\}.
\end{align}

\item $G_a(A_3)=A_3=r^4+u_zr^2+{\bf u .r}$, \\
where ${\bf r}$ and ${\bf u}$ is the position vector to $(r,\theta)$ and $(u_r,u_{\theta})$ surface
of $M$ and $U$, respectively.

For $G_a(A_3)$,
the Maxwell set is same as that of $A_3$
\begin{equation}
\Mset(G_a(A_3))=\{(u_r,u_z,u_{\theta})|u_r=0,u_z<0\}.
\end{equation}
\end{itemize}

\begin{figure}
\begin{center}
\includegraphics[width=11cm,clip]{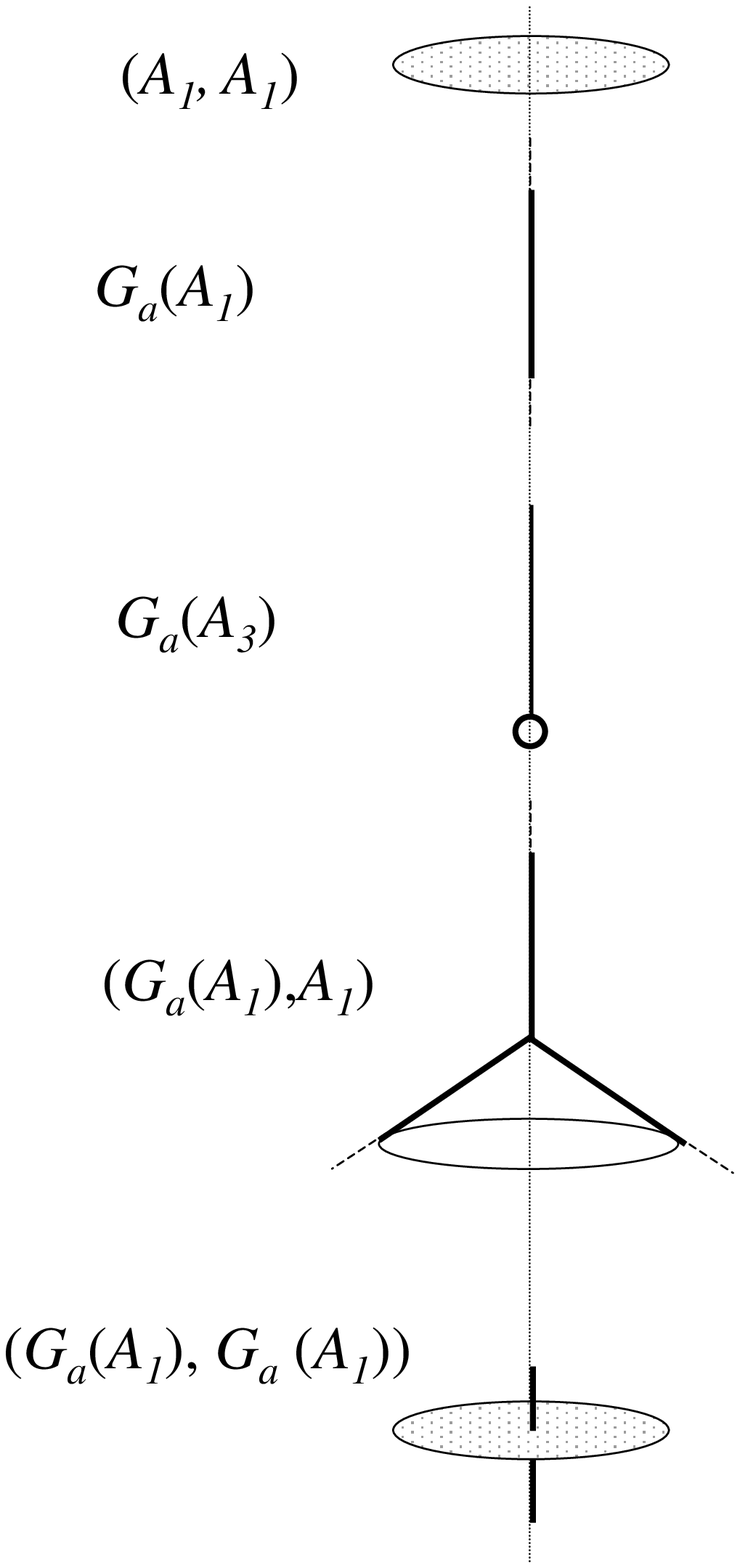} 
\end{center}
\caption{The Maxwell sets in three-dimensional axially symmetric spacetime on its fixed points. 
They are the Maxwell sets of Class D multiunfoldings.}
\label{fig:axcB}
\end{figure}

Their Maxwell sets are illustrated in Fig.\ref{fig:axcB}.

Here it should be noted that the Maxwell set stably possesses one-dimensional
component $G_a(A_1)$, $(G_a(A_1),A_1)$, $(G_a(A_1),G_a(A_1))$ or $G_a(A_3)$,
while only two-dimensional structure is  stable without symmetry\cite{MS2}\cite{SK}.

\begin{figure}
\begin{center}
\includegraphics[width=11cm,clip]{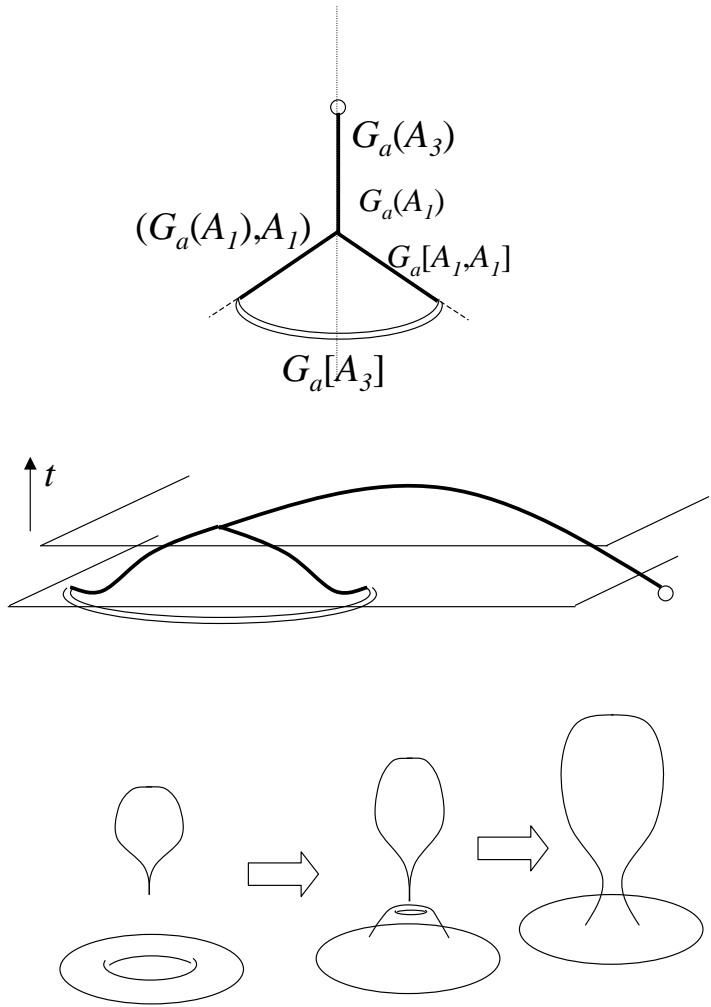} 
\end{center}
\caption{An example of the axially symmetric Maxwell set is shown. That is composed of $G_a(A_3)$, $(G_a(A_1),A_1)$ and
$G_a[A_3]$, which are connected by $G_a(A_1)$ and $G_a[A_1,A_1]$.
Next figure is the crease set which is given by embedding the Maxwell set into a spacetime as a spatial set.
The bottom figure indicate the change of the topology of an event horizon with that crease set.}
\label{fig:axexp}
\end{figure}

From the above classifications, we are able to see the crease set.
As stated in \cite{MS1}\cite{SK}, the crease set is an acausal embedding of
the Maxwell set into the spacetime so that the crease set is homeomorphic to a ball
and approaches to null at its boundary. We give a figure of an example of axially symmetric 
Maxwell set in Fig.\ref{fig:axexp}.

In the example in Fig.\ref{fig:axexp}, by an appropriate timeslicing, it is observed
that toroidal black hole and spherical black hole coalesce.
At the lower spatial hypersurface, the section of the crease set is a ring and a point.
They would be a crease of a toroidal event horizon and a cusp on a spherical event horizon.
As time goes by they approach each other and the ring shrinks, and then they unite and disappear.
This implies the coalescence of toroidal black hole and a spherical black hole.

To find a example in Fig.\ref{fig:axexp}, we have combined the structures of the Maxwell set so that a total
Maxwell set is homeomorphic to a ball, since it is required by the assumption that the event horizon
finally becomes a single sphere. Nevertheless it might be important to consider topologically non-trivial
final state of the event horizon, which becomes meaningful in studies of higher-dimensional black hole.
For example, for a toroidal final event horizon, we compose the Maxwell set so that it is homeomorphic to
a ring.

\section{conclusion and discussions}

We have investigated the Maxwell set under a discrete symmetry and a continuous symmetry.
In the context of singularity theory, the stable Maxwell set is classified.
Then we concretely showed the classification of the stable Maxwell set with 
reflection symmetry in three dimensional spacetime and axial symmetry in four-dimensional spacetime.
Remarkable fact is that by the axial symmetry  one-dimensional segments of the crease set is possible to become 
stable while
it is not stable in four-dimensional spacetime without symmetry.

It might be worth to comment that the present result also is relevant in the case of almost symmetric 
where the symmetry is dynamically stable, if a very fine structure is neglected.
Some cases will be almost axially symmetric in realistic gravitational collapse.
Nevertheless it is not mathematically clear how can we neglect the fine structure.
Under the present situation, our result only suggests the structure of the crease set in almost symmetric spacetime.

By the way, one may be doubtful that peculiar crease sets for example like in Fig.\ref{fig:axexp} can be formed.
However, it will be realistic in the collision of collapsing stars.
In an oblate spheroidal event horizon, it is probable that the crease set is
naturally two dimensional\cite{MS2}\cite{HW}.
If sufficiently oblate spheroidal collapsing star and spherical collapsing star are sufficiently away 
 in direction of their symmetry axis and
if they coalesce, the crease set like Fig.\ref{fig:axexp} will be formed.
Furthermore,
when several black holes coalesce, their crease set will become fairly complicated.

Since these collapses are axially symmetric, it will be possible to simulate them by numerical
calculation of the gravitational collapse.
Not only the fairly simple case in refs.\cite{ST}\cite{NCSA} (these cases are also reflection symmetric
about $x$-$y$ plane) but also some considerably complex
cases should be studied.
With that the various topological black hole will be observed simultaneously.
That will be thought-provoking for the gravitational collapse and phenomena concerning the event horizon, e.g.
quasi-normal mode of gravitational radiation.

\section*{Acknowledgments}
This work is based on another research with Dr. Koike\cite{K-S}.

\flushright

\end{document}